\documentclass[10pt,a4paper,british]{extarticle}
\usepackage[T1]{fontenc}
\usepackage[latin1]{inputenc}
\usepackage{amsmath}
\usepackage{amssymb}
\usepackage{setspace}
\usepackage{esint}
\makeatletter

\@ifundefined{date}{}{\date{}}

\allowdisplaybreaks

\usepackage{babel}

\makeatother

\usepackage{babel}
\begin{document}

\title{Light-Cone Analysis of the Pure Spinor Formalism for the Superstring}

\author{Nathan Berkovits%
\thanks{nberkovi@ift.unesp.br%
}\\
 and\\
 Renann Lipinski Jusinskas%
\thanks{renannlj@ift.unesp.br%
}}

\maketitle

\begin{center}
ICTP South American Institute for Fundamental Research \\
Instituto de F\'isica Te\'orica, UNESP - Univ. Estadual Paulista \\
Rua Dr. Bento T. Ferraz 271, 01140-070, S\~ao Paulo, SP, Brasil.
\par\end{center}

\

\begin{abstract}
Physical states of the superstring can be described in light-cone
gauge by acting with transverse bosonic $\alpha_{-n}^{j}$ and fermionic
$\bar{q}_{-n}^{\dot{a}}$ operators on an $SO\left(8\right)$-covariant
superfield where $j,\dot{a}=1$ to $8$. In the pure spinor formalism,
these states are described in an $SO\left(9,1\right)$-covariant manner
by the cohomology of the BRST charge $Q=\frac{1}{2\pi i}\oint\lambda^{\alpha}d_{\alpha}$.
In this paper, a similarity transformation is found which simplifies
the form of $Q$ and maps the light-cone description of the superstring
vertices into DDF-like operators in the cohomology of $Q$. 
\end{abstract}
\tableofcontents{}

\section{Introduction}

Although the covariant description of string theory is convenient
for amplitude computations and for describing curved backgrounds,
the light-cone description is convenient for computing the physical
spectrum and for proving unitarity. For the manifestly spacetime supersymmetric
string, the light-cone description was worked out over 30 years ago
\cite{Schwarz:1982jn} but the covariant description using the pure
spinor formalism \cite{Berkovits:2000fe} is still being developed.

In this paper, the relation between this covariant and light-cone
superstring descriptions will be analyzed. As in other string theories,
physical states in the pure spinor formalism are covariantly described
by the cohomology of a nilpotent BRST operator. However, because the
pure spinor worldsheet ghost is constrained, evaluation of the BRST
cohomology is not straightforward. By partially solving the pure spinor
constraint, it was proven in \cite{Berkovits:2000nn} that the BRST
cohomology reproduces the correct light-cone spectrum. However, the
proof was complicated and involved an infinite set of ghosts-for-ghosts.
In this paper, the proof will be simplified considerably and an explicit
similarity transformation will be given for mapping light-cone superstring
vertex operators constructed from the $SO\left(8\right)$-covariant
superfields of \cite{Brink:1983pf} into DDF-like vertex operators
in the cohomology of the pure spinor BRST operator.

In bosonic string theory, the covariant BRST operator
\begin{equation}
Q_{B}=-\frac{1}{2\pi i}\oint\left\{ \frac{1}{2}c\partial X^{m}\partial X_{m}+ibc\partial c\right\} ,
\end{equation}
can be mapped by a similarity transformation $R$ to the operator
\begin{equation}
\hat{Q}_{B}=e^{R}Q_{B}e^{-R}=-k^{+}\sum_{n\neq0}c_{-n}\alpha_{n}^{-}+c_{0}\left(\frac{k^{m}k_{m}}{2}+\sum_{n\neq0}\alpha_{-n}^{i}\alpha_{n}^{i}-1\right),
\end{equation}
where $\alpha_{n}^{-}$, $\alpha_{n}^{i}$ and $c_{n}$ are modes
of the $X^{-}$, $X^{i}$ and $c$ variables, and $k^{m}$ is the
momentum, with $\sqrt{2}k^{+}=\left(k^{0}+k^{d-1}\right)$ assumed
to be non-vanishing. Because of the quartet argument, the cohomology
of $\hat{Q}_{B}$ is independent of the $\left(c_{n},b_{n},\alpha_{n}^{+},\alpha_{n}^{-}\right)$
modes, so the physical spectrum is the usual light-cone spectrum constructed
by acting with the transverse $\alpha_{-n}^{i}$ modes on the tachyonic
ground state. Furthermore, the similarity transformation $R$ maps
light-cone vertex operators $V_{\textrm{LC}}$ which only depend on
the transverse variables into the physical DDF vertex operators $V_{DDF}=e^{-R}V_{LC}e^{R}$
which are in the cohomology of $Q_{B}$ \cite{Aisaka:2004ga}.

In the pure spinor formalism, the covariant BRST operator is
\begin{equation}
Q=\frac{1}{2\pi i}\oint\left(\lambda^{\alpha}d_{\alpha}\right),
\end{equation}
where $\alpha=1$ to $16$ is an $SO\left(9,1\right)$ spinor index,
$\lambda^{\alpha}$ is a bosonic spinor ghost satisfying the pure
spinor constraint
\begin{equation}
\lambda^{\alpha}\gamma_{\alpha\beta}^{m}\lambda^{\beta}=0,
\end{equation}
and $d_{\alpha}$ is the Green-Schwarz-Siegel fermionic constraint
which has $8$ first-class and $8$ second-class components. A similarity
transformation $R$ will be found which maps $Q$ into
\begin{equation}
\hat{Q}=e^{R}Qe^{-R}=\frac{1}{2\pi i}\oint\left[\lambda^{a}\left(p_{a}+\frac{\sqrt{2}}{2}\hat{T}\theta_{a}\right)+\bar{\lambda}^{\dot{a}}\left(\bar{p}_{\dot{a}}+\frac{\sqrt{2}}{2}\partial X^{+}\bar{\theta}_{\dot{a}}\right)\right],
\end{equation}
where $\theta^{\alpha}=(\theta^{a},\bar{\theta}^{\dot{a}})$
and $p_{\alpha}=(p_{a},\bar{p}_{\dot{a}})$, and $a$ ($\dot{a}$)
represent the chiral (antichiral) $SO\left(8\right)$ spinor indices.
$\hat{T}$ will include part of the energy-momentum tensor and will
impose the usual Virasoro-like conditions.

The cohomology of $\hat{Q}$ will be argued to consist of states which
are independent of $\left(\theta^{a},p_{a}\right)$ and which are
constructed from the $SO\left(8\right)$-covariant light-cone superfields
$f^{a}(\bar{\theta})e^{ik\cdot X}$ of \cite{Brink:1983pf} by
hitting with the transverse raising operators $\alpha_{-n}^{j}$ (bosonic)
and $\bar{q}_{-n}^{\dot{a}}$ (fermionic) as
\begin{equation}
V_{\textrm{LC}}=\prod_{n,j,\dot{a}}\left(\alpha_{-n}^{j}\vphantom{\bar{q}_{-n}^{\dot{a}}}\right)^{N_{n,j}}\left(\bar{q}_{-n}^{\dot{a}}\vphantom{\alpha_{-n}^{j}}\right)^{N_{n,\dot{a}}}\lambda^{a}f_{a}(\bar{\theta})e^{ik\cdot X}\label{eq:ddfone}
\end{equation}
where $k^{m}k_{m}=-2\sum n\left(N_{n,i}+N_{n,\dot{a}}\right)$. Furthermore,
it will be shown that the similarity transformation $R$ maps the
light-cone vertex operators of \eqref{eq:ddfone} into DDF-like vertex
operators $V_{DDF}=e^{-R}V_{LC}e^{R}$, which are in the cohomology
of the pure spinor BRST operator and which will be described in a
separate paper by one of the authors \cite{Jusinskas:2014vqa}. DDF-like
vertex operators in the pure spinor formalism were first constructed
by Mukhopadhyay \cite{Mukhopadhyay:2005zu} using a Wess-Zumino-like
gauge choice which breaks manifest supersymmetry, whereas the recent
construction of Jusinskas \cite{Jusinskas:2014vqa} uses a supersymmetric
gauge choice which simplifies the analysis and enables an explicit
$SO\left(8\right)$ superfield description of the whole physical spectrum. 

In section \ref{sec:Superparticle}, the superparticle will be discussed
and a similarity transformation $R$ will be constructed which maps
the superparticle BRST operator into a simple quadratic form and maps
the light-cone $SO\left(8\right)$-covariant superfield of \cite{Brink:1983pf}
into the super-Yang-Mills vertex operator in the pure spinor formalism.
In section \ref{sec:Superstring}, this construction will be generalized
to the superstring such that the similarity transformation maps the
light-cone vertex operators of \eqref{eq:ddfone} into the DDF-like
vertex operators of \cite{Jusinskas:2014vqa} in the cohomology of the
pure spinor BRST operator.

\section{Superparticle\label{sec:Superparticle}}

In this section, the pure spinor formulation of the superparticle
will first be reviewed and a similarity transformation will then be
presented which makes the massless constraint explicit in the BRST
operator and maps the super-Yang-Mills vertex operator into the light-cone
$SO\left(8\right)$ superfield of \cite{Brink:1983pf}.

\subsection{Review of the pure spinor superparticle}

The pure spinor superparticle was extensively discussed in \cite{Berkovits:2001rb}
and is described by the first order action
\begin{equation}
S=\int d\tau\left\{ \dot{X}_{m}P^{m}-\frac{1}{2}P^{m}P_{m}+\dot{\lambda}^{\alpha}\omega_{\alpha}-i\dot{\theta}^{\alpha}p_{\alpha}\right\} ,\label{eq:SPaction}
\end{equation}
containing the pure spinor ghost $\lambda^{\alpha}$ and the anti-ghost
$\omega_{\alpha}$. Note that the dot above the fields represent derivatives
with respect to $\tau$.

The BRST charge is defined as
\begin{equation}
Q=\lambda^{\alpha}d_{\alpha},\label{eq:PSparticleBRST}
\end{equation}
where
\begin{equation}
d_{\alpha}=p_{\alpha}-\frac{i}{2}P_{m}\left(\gamma^{m}\theta\right)_{\alpha}
\end{equation}
is the supersymmetric derivative, and the supersymmetry generators
are
\begin{equation}
q_{\alpha}=p_{\alpha}+\frac{i}{2}P_{m}\left(\gamma^{m}\theta\right)_{\alpha}.
\end{equation}

Canonical quantization of \eqref{eq:SPaction} gives\begin{subequations}
\begin{eqnarray}
\left[X^{m},P^{n}\right] & = & i\eta^{mn},\\
\left\{ \theta^{\alpha},p_{\beta}\right\}  & = & i\delta_{\beta}^{\alpha}.
\end{eqnarray}
\end{subequations}Note that $\left\{ q_{\alpha},d_{\beta}\right\} =\left[q_{\alpha},P_{m}\right]=0$
and $\left\{ d_{\alpha},d_{\beta}\right\} =P_{m}\gamma_{\alpha\beta}^{m}$.

To compare this covariant description with the light-cone description,
chiral and antichiral $SO\left(9,1\right)$ spinors will be decomposed
into their $SO\left(8\right)$ components as $\theta^{\alpha}\to(\theta^{a},\bar{\theta}^{\dot{a}})$
and $d_{\alpha}\to(d_{a},\bar{d}_{\dot{a}})$ where $a,\dot{a}=1,\ldots,8$,
and the $SO\left(9,1\right)$ vectors will be decomposed as $X^{m}\to(X^{j},X^{+},X^{-})$
for $j=1,\ldots,8$. The precise conventions of this $SO\left(8\right)$
decomposition are discussed in appendix \ref{sec:conventions}, where
the $SO\left(9,1\right)$ gamma-matrices are expressed in terms of
the $SO(8)$ Pauli matrices $\sigma_{a\dot{a}}^{i}$. In terms of
these Pauli matrices, the $D=10$ pure spinor constraint $\lambda^{\alpha}\gamma_{\alpha\beta}^{m}\lambda^{\beta}=0$
takes the form
\begin{equation}
\begin{array}{ccc}
\lambda_{a}\lambda_{a}=0, & \bar{\lambda}_{\dot{a}}\bar{\lambda}_{\dot{a}}=0, & \lambda^{a}\sigma_{a\dot{a}}^{i}\bar{\lambda}^{\dot{a}}=0.\end{array}
\end{equation}

\subsection{Similarity transformation}

Since the cohomology of \eqref{eq:PSparticleBRST} is described by
the $\mathcal{N}=1$ $D=10$ super Yang-Mills superfield, the BRST
operator must impose the Siegel constraint $P^{m}\left(\gamma^{+}\gamma_{m}d\right)_{a}=0$,
which is the generator of $8$ independent kappa symmetries \cite{Siegel:1983hh}
in the Green-Schwarz formalism. To see that, first note that we can
make the constraint explicit in $Q$ by performing a similarity transformation
generated by
\begin{equation}
R=i\frac{P^{i}\bar{N}_{i}}{P^{+}},\label{eq:similarSP1}
\end{equation}
where
\begin{equation}
\bar{N}^{i}=-\frac{1}{\sqrt{2}}\left(\lambda^{a}\sigma_{a\dot{a}}^{i}\bar{\omega}^{\dot{a}}\right).
\end{equation}
Observe that
\begin{eqnarray}
Q^{'} & \equiv & e^{R}Qe^{-R}\nonumber \\
 & = & \lambda^{a}G_{a}+\bar{\lambda}^{\dot{a}}\bar{d}_{\dot{a}},\label{eq:qprime}
\end{eqnarray}
where
\begin{equation}
G_{a}\equiv\frac{P^{m}}{2P^{+}}(\gamma^{+}\gamma_{m}d)_{a}=d_{a}-\frac{P_{i}\left(\sigma^{i}\bar{d}\right)_{a}}{P^{+}\sqrt{2}}
\end{equation}
and satisfies\begin{subequations}
\begin{eqnarray}
\left\{ G_{a},\bar{d}_{\dot{a}}\right\}  & = & 0,\label{eq:firsteq}\\
\left\{ G_{a},G_{b}\right\}  & = & -\frac{\eta_{ab}}{\sqrt{2}}\left(\frac{P^{2}}{P^{+}}\right),
\end{eqnarray}
\end{subequations}with $P^{2}=-2P^{+}P^{-}+P_{i}P^{i}$. It will
be important to note that equation \eqref{eq:firsteq} implies that
nilpotency of $Q_{p}^{'}$ does not rely any more on the pure spinor
constraint $\lambda^{a}\bar{\lambda}^{\dot{a}}\sigma_{a\dot{a}}^{i}=0$.

The next step in simplifying the BRST operator is to perform the further
similarity transformation generated by
\begin{eqnarray}
\hat{R} & \equiv & i\theta^{a}\left(p_{a}-G_{a}\right)\label{eq:similarSP2}\\
 & = & \frac{i}{\sqrt{2}}\left(\theta^{a}\sigma_{a\dot{a}}^{i}\bar{p}^{\dot{a}}\right)\frac{P_{i}}{P^{+}}.\nonumber 
\end{eqnarray}
Unlike $R$ of \eqref{eq:similarSP1}, $\hat{R}$ does not commute
with the supersymmetry generators and transforms the various operators
as $\hat{\mathcal{O}}\equiv e^{\hat{R}}\mathcal{O}e^{-\hat{R}}$:
\begin{subequations}
\begin{eqnarray}
\hat{\bar{d}}_{\dot{a}} & = & \bar{p}_{\dot{a}}-\frac{i}{\sqrt{2}}P^{+}\bar{\theta}_{\dot{a}},\\
\hat{G}_{a} & = & p_{a}+\frac{i\theta_{a}}{2\sqrt{2}}\left(\frac{P^{2}}{P^{+}}\right),\\
\hat{\bar{q}}_{\dot{a}} & = & \bar{p}_{\dot{a}}+\frac{i}{\sqrt{2}}P^{+}\bar{\theta}_{\dot{a}},\\
\hat{q}_{a} & = & p_{a}+\frac{1}{\sqrt{2}}\left(\sigma^{i}\hat{\bar{q}}\right)_{a}\frac{P_{i}}{P^{+}}-\frac{i\theta_{a}}{2\sqrt{2}}\left(\frac{P^{2}}{P^{+}}\right).
\end{eqnarray}
\end{subequations}As expected, $\hat{\bar{d}}_{\dot{a}}$ and $\hat{G}_{a}$
are supersymmetric with respect to the transformed supersymmetry generators
$\hat{q}_{a}$ and $\hat{\bar{q}}_{\dot{a}}$. After performing this
second similarity transformation, the BRST charge $\hat{Q}\equiv e^{\hat{R}}Q'e^{-\hat{R}}$
takes the simple form
\begin{equation}
\hat{Q}=\lambda^{a}\left(p_{a}+\frac{i}{2\sqrt{2}}\frac{P^{2}}{P^{+}}\theta_{a}\right)+\bar{\lambda}^{\dot{a}}\left(\bar{p}_{\dot{a}}-\frac{i}{\sqrt{2}}P^{+}\bar{\theta}_{\dot{a}}\right),\label{eq:newBRSTSP}
\end{equation}
where the mass-shell constraint $P^{m}P_{m}$ now appears explicitly.

\subsection{Relation with light-cone vertex operators}

Because of the simple form of $\hat{Q}$, it is easy to compute its
cohomology and show equivalence to the light-cone vertex operators.
Consider a state with momentum $k^{m}$ (assuming $k^{+}\neq0$) which
is represented by the ghost-number 1 vertex operator
\begin{equation}
\hat{U}=\lambda^{a}\hat{A}_{a}+\bar{\lambda}^{\dot{a}}\hat{\bar{A}}_{\dot{a}}.
\end{equation}
The $\bar{\lambda}_{\dot{a}}\bar{\lambda}_{\dot{b}}$ component of
$\hat{Q}\hat{U}=0$ implies
\begin{equation}
\hat{\bar{D}}_{\dot{a}}\hat{\bar{A}}_{\dot{b}}+\hat{\bar{D}}_{\dot{b}}\hat{\bar{A}}_{\dot{a}}=\eta_{\dot{a}\dot{b}}\Omega
\end{equation}
for some superfield $\Omega$, where $\hat{\bar{D}}_{\dot{a}}=\bar{\partial}_{\dot{a}}-\frac{k^{+}}{\sqrt{2}}\bar{\theta}_{\dot{a}}$.
The above equation implies that $\hat{\bar{A}}_{\dot{a}}=-\frac{1}{\sqrt{2}k^{+}}\hat{\bar{D}}_{\dot{a}}\Omega$,
which can be set to zero by the gauge transformation $\delta\hat{A}_{\alpha}=-\frac{1}{\sqrt{2}k^{+}}\hat{D}_{\alpha}\Omega$.

In the gauge $\hat{\bar{A}}_{\dot{a}}=0$, the $\lambda_{a}\bar{\lambda}_{\dot{a}}$
component of $\hat{Q}\hat{U}=0$ (together with the constraint $\lambda^{a}\bar{\lambda}^{\dot{a}}\sigma_{a\dot{a}}^{i}=0$)
implies that
\begin{equation}
\hat{\bar{D}}_{\dot{a}}\hat{A}_{a}=\hat{A}_{i}\sigma_{a\dot{a}}^{i},\label{eq:SO8constraint}
\end{equation}
for some superfield $\hat{A}_{i}$. Equation \eqref{eq:SO8constraint}
is precisely the constraint on the $SO\left(8\right)$-covariant superfield
described in \cite{Brink:1983pf}, and the most general solution is
\begin{equation}
\hat{A}_{a}=\Phi\left(\theta\right)f_{a}(\bar{\theta})e^{ik\cdot X},
\end{equation}
where $\Phi(\theta)$ is a generic scalar function of $\theta_{a}$
and, as shown in \cite{Brink:1983pf}, $f_{a}(\bar{\theta})$ is a
light-cone super-Yang-Mills superfield depending on an $SO\left(8\right)$
vector $a^{j}$ and an $SO\left(8\right)$ chiral spinor $\chi^{a}$,
denoting the transverse polarizations of the gluon and gluino. The
explicit formula for $f_{a}(\bar{\theta})$ is
\begin{eqnarray}
f_{a}(\bar{\theta}) & = & a^{i}\left(\sigma^{l}\bar{\theta}\right)_{a}\left\{ \eta_{il}+\left(\frac{1}{3!}\right)\tilde{\theta}_{il}+\left(\frac{1}{5!}\right)\tilde{\theta}_{ij}\tilde{\theta}_{jl}+\left(\frac{1}{7!}\right)\tilde{\theta}_{ij}\tilde{\theta}_{jk}\tilde{\theta}_{kl}\right\} +\left(\frac{\sqrt{2}}{k^{+}}\right)\chi_{a}\label{eq:SO8spinorsolution}\\
 &  & +\left(\chi\sigma^{i}\bar{\theta}\right)\left(\sigma^{l}\bar{\theta}\right)_{a}\left\{ \left(\frac{1}{2!}\right)\eta_{il}+\left(\frac{1}{4!}\right)\tilde{\theta}_{il}+\left(\frac{1}{6!}\right)\tilde{\theta}_{ij}\tilde{\theta}_{jl}+\left(\frac{1}{8!}\right)\tilde{\theta}_{ij}\tilde{\theta}_{jk}\tilde{\theta}_{kl}\right\} \nonumber 
\end{eqnarray}
where $\tilde{\theta}_{ij}\equiv\left(\frac{k^{+}}{\sqrt{2}}\right)\bar{\theta}_{\dot{a}}\bar{\theta}_{\dot{c}}\sigma_{\dot{a}\dot{c}}^{ij}$.

Finally, the $\lambda^{a}\lambda^{b}$ component of $\hat{Q}\hat{U}=0$
implies that
\begin{equation}
f_{a}(\bar{\theta})\hat{D}_{b}\Phi(\theta)+f_{b}(\bar{\theta})\hat{D}_{a}\Phi(\theta)=\delta_{ab}\Sigma\label{eq:phieq}
\end{equation}
for some superfield $\Sigma$ where
\begin{equation}
\hat{D}_{b}\equiv\frac{\partial}{\partial\theta^{b}}+\frac{\sqrt{2}}{4}\left(\frac{k^{m}k_{m}}{k^{+}}\right)\theta_{b}.
\end{equation}
Equation \eqref{eq:phieq} can only be satisfied if $\hat{D}_{a}\Phi=0$.
Since $\hat{D}_{a}\hat{D}_{a}=2\sqrt{2}\left(\frac{k^{m}k_{m}}{k^{+}}\right)$,
$\hat{D}_{a}\Phi=0$ implies both that $k^{2}=0$ and that $\frac{\partial}{\partial\theta^{a}}\Phi=0$
(\emph{i.e.} $\Phi$ is constant). After rescaling the polarizations
by the constant $\Phi$, one finally obtains that the states in the
ghost-number one cohomology of $\hat{Q}$ are described by
\begin{equation}
\hat{U}=\lambda^{a}f_{a}(\bar{\theta})e^{ik\cdot X},\label{eq:spu}
\end{equation}
where $f^{a}(\bar{\theta})$ is the $SO\left(8\right)$-covariant
light-cone superfield of \eqref{eq:SO8spinorsolution} and $k^{m}k_{m}=0$.

States in the cohomology of the original pure spinor BRST operator
are directly obtained from $\hat{U}$ by defining
\begin{equation}
U\equiv e^{-\left(R+\hat{R}\right)}\hat{U}e^{\left(R+\hat{R}\right)}=\lambda^{a}f_{a}(\hat{\theta})e^{ik\cdot X},
\end{equation}
where $\hat{\theta}_{\dot{a}}\equiv\bar{\theta}_{\dot{a}}+\frac{k^{i}}{\sqrt{2}k^{+}}\left(\sigma^{i}\theta\right)_{\dot{a}}$.
Note that when the state has vanishing transverse momentum, $k^{i}=0$,
the similarity transformations $R$ and $\hat{R}$ of \eqref{eq:similarSP1}
and \eqref{eq:similarSP2} vanish and
\begin{equation}
U|_{k^{i}=0}=\hat{U}=\lambda^{a}f_{a}(\bar{\theta})e^{-ik^{+}X^{-}}.\label{Unormal}
\end{equation}
It is easy to verify that $U$ is the usual vertex operator $\lambda^{\alpha}A_{\alpha}(X,\theta)$
in the gauge $(\gamma^{+}A)_{\dot{a}}=0$.

Now we will proceed to the more intricate case of the superstring.

\section{Superstring\label{sec:Superstring}}

In this section, we will repeat the analysis done for the superparticle.
After reviewing the pure spinor description of the superstring, we
will show that the pure spinor BRST charge $Q=\frac{1}{2\pi i}\oint\left(\lambda^{\alpha}d_{\alpha}\right)$
can be written after a similarity transformation as
\begin{equation}
\hat{Q}=\frac{1}{2\pi i}\oint\left[\lambda^{a}\left(p_{a}+\frac{\sqrt{2}}{2}\hat{T}\theta_{a}\right)+\bar{\lambda}^{\dot{a}}\left(\bar{p}_{\dot{a}}+\frac{\sqrt{2}}{2}\partial X^{+}\bar{\theta}_{\dot{a}}\right)\right],\label{eq:newPSBRST}
\end{equation}
where
\begin{eqnarray}
\hat{T} & \equiv & -\frac{1}{2}\left(\frac{\partial X^{m}\partial X_{m}}{\partial X^{+}}\right)+i\left(\frac{\bar{p}_{\dot{a}}\partial\bar{\theta}_{\dot{a}}}{\partial X^{+}}\right)+\frac{1}{2}\partial\left(\frac{\hat{J}}{\partial X^{+}}\right)\nonumber \\
 &  & -\left(\frac{i\sqrt{2}}{4}\right)\frac{\hat{\bar{d}}_{\dot{a}}\partial\hat{\bar{d}}_{\dot{a}}}{\left(\partial X^{+}\right)^{2}}-\left(\frac{1}{2}\right)\frac{\left(\partial\ln\left(\partial X^{+}\right)\right)^{2}}{\partial X^{+}}\label{eq:That}
\end{eqnarray}
and
\begin{eqnarray}
\hat{J} & \equiv & -\bar{\omega}_{\dot{a}}\bar{\lambda}_{\dot{a}},\label{eq:ghostnumberdot}\\
\hat{\bar{d}}_{\dot{a}} & \equiv & \bar{p}_{\dot{a}}+\frac{\sqrt{2}}{2}\partial X^{+}\bar{\theta}_{\dot{a}}.
\end{eqnarray}

The structure of $\hat{Q}$ for the superstring closely resembles
\eqref{eq:newBRSTSP} for the superparticle, and one can verify that
$\hat{Q}$ is nilpotent using the OPE's
\begin{eqnarray}
\hat{T}\left(z\right)\bar{\lambda}_{\dot{a}}\hat{\bar{d}}_{\dot{a}}\left(y\right) & \sim & \textrm{regular},\\
\hat{T}\left(z\right)\hat{T}\left(y\right) & \sim & \textrm{regular}.
\end{eqnarray}
The cohomology of $\hat{Q}$ will be shown to reproduce the usual
light-cone superstring spectrum where the similarity transformation
maps the DDF-like vertex operators of \cite{Jusinskas:2014vqa} into
the $8$ transverse bosonic and fermionic operators, $\alpha_{-n}^{j}$
and $\bar{q}_{-n}^{\dot{a}}$, which create massive superstring states
from the massless ground state in light-cone gauge.

\subsection{Review of the pure spinor superstring}

The matter (holomorphic) sector of the pure spinor formalism is constructed
from the Green-Schwarz-Siegel variables of \cite{Siegel:1985xj} and
is described by the free action
\begin{equation}
S_{\textrm{matter}}=\frac{1}{2\pi}\int d^{2}z\left(\frac{1}{2}\partial X^{m}\bar{\partial}X_{m}+ip_{\beta}\bar{\partial}\theta^{\beta}\right),
\end{equation}
and the free field OPE's\begin{subequations}
\begin{eqnarray}
X^{m}\left(z,\bar{z}\right)X^{n}\left(y,\bar{y}\right) & \sim & -\eta^{mn}\ln\left|z-y\right|^{2},\\
p_{\alpha}\left(z\right)\theta^{\beta}\left(y\right) & \sim & \frac{i\delta_{\alpha}^{\beta}}{z-y}.
\end{eqnarray}
\end{subequations}

The supersymmetry charge is
\begin{equation}
q_{\alpha}=\frac{1}{2\pi}\oint\left\{ -p_{\alpha}+\frac{1}{2}\partial X^{m}\left(\gamma_{m}\theta\right)_{\alpha}+\frac{i}{24}\left(\theta\gamma^{m}\partial\theta\right)\left(\gamma_{m}\theta\right)_{\alpha}\right\} ,
\end{equation}
satisfying $\left\{ q_{\alpha},q_{\beta}\right\} =P_{m}\gamma_{\alpha\beta}^{m}$,
where $P^{m}\equiv\frac{1}{2\pi}\oint\partial X^{m}.$ The usual supersymmetric
invariants are
\begin{eqnarray}
\Pi^{m} & = & \partial X^{m}+\frac{i}{2}\left(\theta\gamma^{m}\partial\theta\right),\\
d_{\alpha} & = & p_{\alpha}+\frac{1}{2}\partial X^{m}\left(\gamma_{m}\theta\right)_{\alpha}+\frac{i}{8}\left(\theta\gamma^{m}\partial\theta\right)\left(\gamma_{m}\theta\right)_{\alpha},
\end{eqnarray}
and the OPE's among them are easily computed to be\begin{subequations}
\begin{eqnarray}
\Pi^{m}\left(z\right)\Pi^{n}\left(y\right) & \sim & -\frac{\eta^{mn}}{\left(z-y\right)^{2}},\\
d_{\alpha}\left(z\right)\Pi^{m}\left(y\right) & \sim & -\frac{\gamma_{\alpha\beta}^{m}\partial\theta^{\beta}}{\left(z-y\right)},\\
d_{\alpha}\left(z\right)d_{\beta}\left(y\right) & \sim & i\frac{\gamma_{\alpha\beta}^{m}\Pi_{m}}{\left(z-y\right)}.
\end{eqnarray}
\end{subequations}

The main feature of the formalism is its simple BRST charge, given
by
\begin{equation}
Q=\frac{1}{2\pi i}\oint\left(\lambda^{\alpha}d_{\alpha}\right),\label{eq:PSBRST}
\end{equation}
where $\lambda^{\alpha}$ is the bosonic ghost. Nilpotency of $Q$
is achieved when $\lambda^{\alpha}$ is constrained by $\lambda\gamma^{m}\lambda=0$,
the pure spinor condition. The conjugate of $\lambda^{\alpha}$ will
be represented by $\omega_{\alpha}$ and they can be described by
the Lorentz covariant action
\begin{equation}
S_{\lambda}=\frac{1}{2\pi}\int d^{2}z\left(\omega_{\alpha}\bar{\partial}\lambda^{\alpha}\right),
\end{equation}
which has the gauge invariance $\delta\omega_{\alpha}=\phi_{m}\left(\gamma^{m}\lambda\right)_{\alpha}$
due to the pure spinor constraint. The gauge invariant quantities
are the Lorentz current, $N^{mn}=-\frac{1}{2}\left(\omega\gamma^{mn}\lambda\right)$,
the ghost number current, $J=-\omega\lambda$, and the energy-momentum
tensor of the ghost sector, $T_{\lambda}=-\omega\partial\lambda$.
The pure spinor constraint also implies the classical constraints
on the currents:
\begin{eqnarray}
N^{mn}\left(\gamma_{n}\lambda\right)_{\alpha}+\frac{1}{2}J\left(\gamma^{m}\lambda\right)_{\alpha} & = & 0.\label{eq:PScurrentsconstraints}\\
N_{mn}\left(\gamma^{mn}\partial\lambda\right)^{\alpha}+J\partial\lambda^{\alpha} & = & -4\lambda^{\alpha}T_{\lambda}.
\end{eqnarray}

The physical open string spectrum is described by the ghost number
one cohomology of $Q$. For example, the massless states are described
by the unintegrated vertex
\begin{equation}
U=\lambda^{\alpha}A_{\alpha}\left(X,\theta\right),\label{eq:PSmasslessU}
\end{equation}
where $A_{\alpha}$ is a superfield composed of the zero-modes of
$\left(X^{m},\theta^{\alpha}\right)$. Note that
\begin{eqnarray}
\left\{ Q,U\right\}  & = & \lambda^{\alpha}\lambda^{\beta}D_{\alpha}A_{\beta},
\end{eqnarray}
where
\begin{equation}
D_{\alpha}\equiv i\partial_{\alpha}-\frac{1}{2}\left(\gamma^{m}\theta\right)_{\alpha}\partial_{m},\label{eq:defsuperderivative}
\end{equation}
with $\partial_{\alpha}=\frac{\partial}{\partial\theta^{\alpha}}$,
$\partial_{m}=\frac{\partial}{\partial X^{m}}$. Since $\lambda^{\alpha}$
is a pure spinor, $\lambda^{\alpha}\lambda^{\beta}\propto\gamma_{mnpqr}^{\alpha\beta}\left(\lambda\gamma^{mnpqr}\lambda\right)$
and $\left\{ Q,U\right\} =0$ implies the linearized super Yang-Mills
equation of motion \cite{Howe:1991mf}:
\begin{equation}
D\gamma^{mnpqr}A=0.\label{eq:eomSYM}
\end{equation}

The integrated version of \eqref{eq:PSmasslessU} is given by
\begin{equation}
V=\frac{1}{2\pi i}\oint\left\{ \Pi^{m}A_{m}+i\partial\theta^{\alpha}A_{\alpha}+id_{\alpha}W^{\alpha}+N^{mn}F_{mn}\right\} ,\label{eq:PSmasslessI}
\end{equation}
where $A^{m}$ and $W^{\alpha}$ are the super Yang-Mills fields,
constrained by\begin{subequations}\label{eq:superfields}
\begin{eqnarray}
A_{m} & \equiv & \frac{1}{8i}\left(D_{\alpha}\gamma_{m}^{\alpha\beta}A_{\beta}\right),\label{eq:supervector}\\
\left(\gamma_{m}W\right)_{\alpha} & \equiv & \left(D_{\alpha}A_{m}+\partial_{m}A_{\alpha}\right),\label{eq:superspinor}
\end{eqnarray}
\end{subequations}and $F_{mn}=\frac{1}{2}\left(\partial_{m}A_{n}-\partial_{n}A_{m}\right)$.

\subsection{Similarity transformation}

To show that the cohomology of \eqref{eq:PSBRST} describes the light-cone
superstring spectrum, it will be convenient to follow the same procedure
as in the previous section for the superparticle. The superstring
version of the similarity transformation of \eqref{eq:similarSP1}
is 
\begin{equation}
R=-\frac{1}{2\pi i}\oint\left\{ \frac{\bar{N}_{i}\Pi^{i}}{\Pi^{+}}\right\} ,\label{eq:similarity1}
\end{equation}
and transforms $\lambda^{\alpha}$ and $d_{\alpha}$ as
\begin{eqnarray*}
e^{R}\lambda_{a}e^{-R} & = & \lambda_{a},\\
e^{R}\bar{\lambda}_{\dot{a}}e^{-R} & = & \bar{\lambda}_{\dot{a}}-\frac{\left(\sigma^{i}\lambda\right)_{\dot{a}}\Pi_{i}}{\sqrt{2}\Pi^{+}}-\left(\frac{\sqrt{2}}{4}\right)\frac{\left(\sigma^{i}\lambda\right)_{\dot{a}}\partial\bar{N}^{i}}{\left(\Pi^{+}\right)^{2}},\\
e^{R}d_{a}e^{-R} & = & d_{a}-\frac{\left(\sigma^{i}\partial\bar{\theta}\right)_{a}\bar{N}_{i}}{\Pi^{+}}-\sqrt{2}\frac{\partial\theta_{a}\bar{N}_{i}\Pi^{i}}{\left(\Pi^{+}\right)^{2}},\\
e^{R}\bar{d}_{\dot{a}}e^{-R} & = & \bar{d}_{\dot{a}}-\frac{\left(\sigma^{i}\partial\theta\right)_{\dot{a}}\bar{N}_{i}}{\Pi^{+}}.
\end{eqnarray*}
Using the above relations together with the properties $\bar{N}_{i}\left(\sigma^{i}\bar{\lambda}\right)_{a}=\sqrt{2}\lambda_{a}\hat{J}$
and $\bar{N}_{i}\bar{N}^{i}=0$, which follow from the $SO\left(8\right)$
decomposition of \eqref{eq:PScurrentsconstraints}, we obtain
\begin{eqnarray}
Q^{'} & \equiv & \frac{1}{2\pi i}\oint e^{R}\left(\lambda^{\alpha}d_{\alpha}\right)e^{-R}\\
 & = & \frac{1}{2\pi i}\oint\left\{ \lambda^{a}\left(G_{a}-\sqrt{2}\frac{\partial\theta_{a}}{\Pi^{+}}\hat{J}\right)+\bar{\lambda}^{\dot{a}}\bar{d}_{\dot{a}}-\left(\frac{\sqrt{2}}{4}\right)\frac{\left(\lambda^{a}\sigma_{a\dot{a}}^{i}\bar{d}^{\dot{a}}\right)\partial\bar{N}^{i}}{\left(\Pi^{+}\right)^{2}}\right\} ,\label{eq:BRSTp}
\end{eqnarray}
where
\begin{equation}
G_{a}\equiv d_{a}-\frac{\left(\sigma^{i}\bar{d}\right)_{a}}{\sqrt{2}}\left(\frac{\Pi^{i}}{\Pi^{+}}\right).\label{eq:defGa}
\end{equation}
Note that although normal-ordering contributions are being ignored
in the explicit computations, the only terms that can receive quantum
corrections are
\begin{eqnarray}
\frac{\partial^{2}\theta_{a}}{\Pi^{+}} & \textrm{and} & \partial\left(\frac{\partial\theta_{a}}{\Pi^{+}}\right)\label{eq:defGb}
\end{eqnarray}
 and their coefficients can be determined by requiring nilpotency
of $Q^{'}$.

The term proportional to $\partial\bar{N}^{i}$ in \eqref{eq:BRSTp}
did not appear in the superparticle BRST operator of \eqref{eq:qprime},
however, it can fortunately be removed by performing a second similarity
transformation generated by 
\begin{equation}
R'=-\frac{\sqrt{2}}{8\pi}\oint\left\{ \frac{\bar{N}_{i}\left(\partial\theta\sigma^{i}\bar{d}\right)}{(\Pi^{+})^{2}}\right\} .\label{eq:similarity2}
\end{equation}
After this transformation, the BRST operator $Q''=e^{R'}Q'e^{-R'}$
is 
\begin{eqnarray}
Q^{''} & = & \frac{1}{2\pi i}\oint\left\{ \lambda^{a}\left(G_{a}+\partial\theta_{b}H_{ab}-\left(\frac{\sqrt{2}}{2}\right)\frac{\partial\theta_{a}\hat{J}}{\Pi^{+}}\right)+\bar{\lambda}_{\dot{a}}\bar{d}_{\dot{a}}\right\} ,\label{eq:BRSTcurrent2}
\end{eqnarray}
where 
\begin{equation}
H_{ab}=-H_{ba}=\left(\frac{i}{4}\right)\frac{\left(\sigma^{i}\bar{d}\right)_{a}\left(\sigma^{i}\bar{d}\right)_{b}}{\left(\Pi^{+}\right)^{2}},
\end{equation}
and the possible normal-ordering contributions have the same form
of those in \eqref{eq:defGb} and can be determined in an analogous
manner. Similar to the superparticle BRST charge $Q^{'}$ of \eqref{eq:qprime},
$Q^{''}$ of \eqref{eq:BRSTcurrent2} is manifestly supersymmetric
and is nilpotent without requiring the pure spinor constraint $\left(\lambda\sigma^{i}\bar{\lambda}\right)=0$
because
\begin{equation}
\left(G_{a}+\partial\theta_{b}H_{ab}-\left(\frac{\sqrt{2}}{2}\right)\frac{\partial\theta_{a}\hat{J}}{\Pi^{+}}\right)\left(z\right)\left(\bar{\lambda}_{\dot{a}}\bar{d}_{\dot{a}}\right)\left(y\right)\sim\textrm{regular}.
\end{equation}

To reduce $Q''$ to the form of $\hat{Q}$ in \eqref{eq:newPSBRST},
one needs to perform a further similarity transformation which is
a generalization of $\hat{R}$ presented in \eqref{eq:similarSP2}
for the superparticle. Expanding in powers of $\theta^{a}$, one finds
that
\begin{eqnarray}
\hat{R} & = & \frac{1}{2\pi i}\oint\left\{ \frac{i}{\sqrt{2}}\frac{\partial X^{i}}{\partial X^{+}}\left(\theta\sigma^{i}\bar{p}\right)+\frac{\sqrt{2}}{8}\frac{\left(\theta\sigma^{i}\bar{p}\right)\left(\partial\theta\sigma^{i}\bar{\theta}\right)}{\partial X^{+}}\right.\nonumber \\
 &  & \left.-\frac{\sqrt{2}}{8}\frac{\left(\theta\sigma^{i}\partial\bar{\theta}\right)\left(\theta\sigma^{i}\hat{\bar{d}}\right)}{\partial X^{+}}-\frac{1}{8}\frac{\left(\partial\theta\sigma^{i}\hat{\bar{d}}\right)\left(\theta\sigma^{i}\hat{\bar{d}}\right)}{\left(\partial X^{+}\right)^{2}}+\frac{i}{2\sqrt{2}}\left(\frac{\theta\partial\theta}{\partial X^{+}}\right)\hat{J}+\ldots\right\} \label{eq:rhat}
\end{eqnarray}
where $...$ denotes terms which are at least cubic order in $\theta^{a}$,
$\theta^{ij}=\theta^{a}\theta^{c}\sigma_{ac}^{ij}$ and
\[
\hat{\bar{d}}_{\dot{a}}\equiv\bar{p}_{\dot{a}}+\frac{\sqrt{2}}{2}\partial X^{+}\bar{\theta}_{\dot{a}}.
\]

The first term in \eqref{eq:rhat} is the same as in the superparticle
$\hat{R}$ of \eqref{eq:similarSP2} while the second is required
to transform the supersymmetry generator $\hat{\bar{q}}_{\dot{a}}\equiv e^{\hat{R}}\bar{q}_{\dot{a}}e^{-\hat{R}}$
to the simple form
\begin{equation}
\hat{\bar{q}}_{\dot{a}}=-\frac{1}{2\pi}\oint\left\{ \bar{p}_{\dot{a}}-\frac{\sqrt{2}}{2}\partial X^{+}\bar{\theta}_{\dot{a}}\right\} .
\end{equation}
The terms in the second line of \eqref{eq:rhat} commute with $\hat{\bar{q}}_{\dot{a}}$
and are necessary so that $\hat{Q}\equiv e^{\hat{R}}Q^{''}e^{-\hat{R}}$
has at most linear dependence on $\theta^{a}$. Using the explicit
terms in \eqref{eq:rhat}, it was verified up to linear order in $\theta^{a}$
that
\begin{equation}
\hat{Q}=\frac{1}{2\pi i}\oint\left[\lambda^{a}\left(p_{a}+\frac{\sqrt{2}}{2}\hat{T}\theta_{a}\right)+\bar{\lambda}^{\dot{a}}\left(\bar{p}_{\dot{a}}+\frac{\sqrt{2}}{2}\partial X^{+}\bar{\theta}_{\dot{a}}\right)\right],\label{eq:newPSBRST2}
\end{equation}
where $\hat{T}$ is defined in \eqref{eq:That}.

\subsection{Relation with light-cone vertex operators}

To compute the cohomology of $\hat{Q}$ of \eqref{eq:newPSBRST2},
note that the zero mode structure of $\hat{Q}$ is the same as in
the superparticle $\hat{Q}$ of \eqref{eq:newBRSTSP}, so the superstring
ground state describing the massless states is 
\begin{equation}
\hat{U}=\lambda^{a}f_{a}(\bar{\theta})e^{ik\cdot X}\label{eq:spu2}
\end{equation}
of \eqref{eq:spu} where $f^{a}(\bar{\theta})$ is the $SO\left(8\right)$-covariant
light-cone superfield of \eqref{eq:SO8spinorsolution} and $k^{m}k_{m}=0$.

To construct massive states in the cohomology, first note that the
integrated vertex operators\begin{subequations}\label{eq:osc} 
\begin{eqnarray}
\alpha_{n}^{j} & \equiv & \frac{1}{2\pi i}\oint\partial X^{j}\exp\left(\frac{in}{k^{+}}X^{+}\right),\\
\bar{q}_{n}^{\dot{a}} & \equiv & -\frac{1}{2\pi}\oint\left(\bar{p}^{\dot{a}}-\frac{\sqrt{2}}{2}\partial X^{+}\bar{\theta}^{\dot{a}}\right)\exp\left(\frac{in}{k^{+}}X^{+}\right),
\end{eqnarray}
\end{subequations}are in the cohomology of $\hat{Q}$ for any value
of $n$. The relation to the usual Laurent modes becomes clear when
$X^{+}\left(z\right)=-ik^{+}\ln\left(z\right)$, \emph{i.e.} in light-cone
gauge where $X^{+}$ is the worldsheet time coordinate. In this gauge,
$\exp\left(\frac{in}{k^{+}}X^{+}\right)=z^{n}$ and we recover the
usual Laurent expansion.

One interpretation of the integrated vertex operators of \eqref{eq:osc}
is as massless integrated vertex operators for the $8$ physical polarizations
of the gluon and gluino with momenta $p^{j}=p^{+}=0$ and $p^{-}=\frac{n}{k^{+}}$.
However, as will be discussed in \cite{Jusinskas:2014vqa}, another
interpretation of \eqref{eq:osc} is as DDF-like operators which act
on the ground state vertex operator of \eqref{eq:spu2} with $k^{i}=0$
to create excited state vertex operators that describe the massive
superstring states. If $X^{+}$ in \eqref{eq:osc} is treated as a
holomorphic variable with the OPE $X^{+}(z)X^{-}(y)\sim\ln\left(z-y\right)$
and $n$ is a positive integer, the contour integral of $\alpha_{-n}^{j}$
and $\bar{q}_{-n}^{\dot{a}}$ around the ground state vertex operator
$\hat{U}=\lambda^{a}f_{a}(\bar{\theta})e^{ik\cdot X}$ will produce
the excited state vertex operators\begin{subequations} 
\begin{eqnarray}
\alpha_{-n}^{j}\hat{U} & = & \frac{1}{\left(n-1\right)!}\left(\partial^{n}X^{j}+\ldots\right)\lambda^{a}f_{a}(\bar{\theta})e^{i\left(k^{j}X^{j}-k^{+}X^{-}-\left(k^{-}+\frac{n}{k^{+}}\right)X^{+}\right)},\\
\bar{q}_{-n}^{\dot{a}}\hat{U} & = & \frac{1}{\left(n-1\right)!}\left[\partial^{n-1}\left(\bar{p}^{\dot{a}}+\frac{ik^{+}}{n\sqrt{2}}\partial\bar{\theta}^{\dot{a}}\right)+\ldots\right]\lambda^{a}f_{a}(\bar{\theta})e{}^{i\left(k^{j}X^{j}-k^{+}X^{-}-\left(k^{-}+\frac{n}{k^{+}}\right)X^{+}\right)},
\end{eqnarray}
\end{subequations}where $\ldots$ denotes terms proportional to derivatives
of $X^{+}$. One can similarly act with any number of $\alpha_{-n}^{j}$
and $\bar{q}_{-n}^{\dot{a}}$ operators on the ground state vertex
operator to construct the general excited state vertex operator
\begin{equation}
\prod_{n>0}\prod_{\dot{a}}\prod_{j}\left(\alpha_{-n}^{j}\vphantom{\bar{q}_{-n}^{\dot{a}}}\right)^{N_{n,j}}\left(\bar{q}_{-n}^{\dot{a}}\vphantom{\alpha_{-n}^{j}}\right)^{N_{n,\dot{a}}}\hat{U}.\label{eq:oscex3}
\end{equation}

Since $\alpha_{-n}^{j}$ and $\bar{q}_{-n}^{\dot{a}}$ commute with
$\hat{Q}$, it is clear that the vertex operators of \eqref{eq:oscex3}
are BRST-closed. And it is easy to see they are not BRST-exact since
the worldsheet variables $\partial X^{j}$ and $\left(\bar{p}_{\dot{a}}-\frac{\sqrt{2}}{2}\partial X^{+}\bar{\theta}_{\dot{a}}\right)$
only appear in $\hat{Q}$ through $\hat{T}$. Furthermore, one expects
that there are no other states in the cohomology of $\hat{Q}$ as
the terms $\left(\lambda^{a}p_{a}\right)$ and $\left(\bar{\lambda}^{\dot{a}}\hat{\bar{d}}_{\dot{a}}\right)$
in \eqref{eq:newPSBRST2} imply that the cohomology is independent
of $\left(\theta^{a},p_{a}\right)$, and depends on $\left(\bar{\theta}_{\dot{a}},\bar{p}_{\dot{a}}\right)$
only through combinations that anticommute with $\hat{\bar{d}}_{\dot{a}}$.
Also, as in bosonic string theory, the dependence on $\left(X^{+},X^{-}\right)$
is completely fixed by $\hat{T}$. So the cohomology of $\hat{Q}$
is expected to be described by the states of \eqref{eq:oscex3} which
are in one-to-one correspondence with the usual light-cone Green-Schwarz
states of the superstring spectrum.

Since the original pure spinor BRST operator $Q$ is related to $\hat{Q}$
by $Q=e^{-R-R^{'}}e^{-\hat{R}}\hat{Q}e^{\hat{R}}e^{R+R^{'}}$, where
the similarity transformations $R$, $R^{'}$ and $\hat{R}$ are defined
in \eqref{eq:similarity1}, \eqref{eq:similarity2} and \eqref{eq:rhat},
covariant BRST-invariant vertex operators in the pure spinor formalism
can be related to the vertex operators of \eqref{eq:oscex3} by acting
with these same similarity transformations.

To see how this works, it will be useful to interpret $\alpha_{n}^{j}$
and $\bar{q}_{n}^{\dot{a}}$ of \eqref{eq:osc} as integrated vertex
operators for a massless gluon and gluino with momenta $p^{j}=p^{+}=0$
and $p^{-}=\frac{n}{k^{+}}$, and to combine them into a super-Yang-Mills
vertex operator by contracting them with the gluon and gluino polarization
$a_{j}$ and $\bar{\chi}_{\dot{a}}$ as
\begin{equation}
\hat{V}_{-n}\equiv a_{j}\alpha_{-n}^{j}-i\bar{\chi}_{\dot{a}}\bar{q}_{-n}^{\dot{a}}.\label{eq:vv}
\end{equation}
After performing the similarity transformation with $\hat{R}$ of
\eqref{eq:rhat}, one finds (up to terms quadratic in $\theta^{a}$)
that
\begin{equation}
V_{-n}^{'}\equiv e^{-\hat{R}}\hat{V}_{-n}e^{\hat{R}}=\frac{1}{2\pi i}\oint\left\{ \Pi_{i}A^{i}+\left(i\partial\bar{\theta}_{\dot{a}}+\frac{n}{\sqrt{2}k^{+}}\bar{d}_{\dot{a}}\right)A^{\dot{a}}\right\} \label{eq:R-PSDDF}
\end{equation}
where $D_{a}A_{\dot{a}}=i\sigma_{a\dot{a}}^{j}A_{j}$, with $D_{a}=\partial_{a}-\frac{n}{k^{+}\sqrt{2}}\theta^{a}$.
$A_{\dot{a}}$ is an $SO\left(8\right)$-superfield that depends only
on $\theta_{a}$ and $X^{+}$, in an exact parallel to $f_{a}\left(\bar{\theta}\right)e^{-ik^{+}X^{-}}$
which depends only on $\bar{\theta}_{\dot{a}}$ and $X^{-}$ and satisfies
the constraint \eqref{eq:SO8constraint}. See \cite{Jusinskas:2014vqa}
for further details on $A_{\dot{a}}$ and how it emerges in the pure
spinor cohomology.

Finally, the gauge fixed version of the integrated massless vertex
of \eqref{eq:PSmasslessI} is obtained by acting with the similarity
transformation $R+R'$ of \eqref{eq:similarity1} and \eqref{eq:similarity2}
which transforms $V'_{-n}$ into
\begin{eqnarray}
V_{-n} & \equiv & e^{-R-R^{'}}V_{-n}^{'}e^{R+R^{'}}\label{eq:R-PSDDF2}\\
 & = & \frac{1}{2\pi i}\oint\left\{ \left(\Pi_{i}-i\frac{n}{k^{+}}\bar{N}_{i}\right)A^{i}+\left(i\partial\bar{\theta}_{\dot{a}}+\frac{n}{\sqrt{2}k^{+}}\bar{d}_{\dot{a}}\right)A_{\dot{a}}\right\} .\nonumber 
\end{eqnarray}
It is straightforward to see that $R^{'}$ commutes with $V_{-n}^{'}$,
and that $R$ is responsible for reintroducing the ghost Lorentz current
$\bar{N}_{i}$ in the vertex.

A detailed discussion of the properties of the DDF-like operators
\eqref{eq:R-PSDDF2} is presented in \cite{Jusinskas:2014vqa}. As shown
there, the superstring spectrum is obtained by acting with the above
operators on the $SO\left(8\right)$-covariant ground state $\hat{U}|_{k^{i}=0}$
of \eqref{eq:spu2}, allowing a systematic description of all massive
pure spinor vertex operators in terms of $SO\left(8\right)$ superfields.

\section{Conclusion}

In this paper, the pure spinor BRST operator $Q=\frac{1}{2\pi i}\oint\left(\lambda^{\alpha}d_{\alpha}\right)$
was mapped by the similarity transformations of $R$, $R^{'}$ and
$\hat{R}$ of \eqref{eq:similarity1}, \eqref{eq:similarity2} and
\eqref{eq:rhat} into the nilpotent operator
\begin{equation}
\hat{Q}=\frac{1}{2\pi i}\oint\left[\lambda^{a}\left(p_{a}+\frac{\sqrt{2}}{2}\hat{T}\theta_{a}\right)+\bar{\lambda}^{\dot{a}}\hat{\bar{d}}_{\dot{a}}\right],\label{eq:newPSBRST3}
\end{equation}
where $\hat{T}$ is defined in \eqref{eq:That}. The cohomology of
$\hat{Q}$ is the usual light-cone Green-Schwarz superstring spectrum
and is described by the vertex operators
\begin{equation}
\prod_{n>0}\prod_{\dot{a}}\prod_{j}\left[\partial^{n-1}\left(\bar{p}^{\dot{a}}+\frac{ik^{+}}{n\sqrt{2}}\partial\bar{\theta}^{\dot{a}}\right)+\ldots\right]^{N_{n,\dot{a}}}\left(\partial^{n}X^{j}+\ldots\right)^{N_{n,j}}\lambda^{a}f_{a}(\bar{\theta})e^{i\left(k^{j}X^{j}-k^{+}X^{-}-\tilde{k}^{-}X^{+}\right)},\label{eq:vo}
\end{equation}
where
\begin{equation}
\tilde{k}^{-}=\frac{1}{k^{+}}\left[\frac{k^{i}k_{i}}{2}+\sum_{n}n\left(N_{n,\dot{a}}+N_{n,j}\right)\right],
\end{equation}
$f_{a}(\bar{\theta})$ is the $SO\left(8\right)$-covariant superfield
\eqref{eq:SO8spinorsolution} of reference \cite{Brink:1983pf}, and
$...$ involves derivatives of $X^{+}$. Finally, the similarity transformations
of $R$, $R'$ and $\hat{R}$ were argued to map the vertex operators
of \eqref{eq:vo} into pure spinor BRST-invariant vertex operators
constructed using the DDF-operators described in \cite{Jusinskas:2014vqa}.

\section*{Acknowledgements}

NB would like to thank CNPq grant 300256/94-9 and FAPESP grants 2009/50639-2
and 2011/11973-4 for partial financial support, and RLJ would like
to thank FAPESP grant 2009/17516-4 for financial support.

\appendix

\section{$SO\left(9,1\right)$ to $SO\left(8\right)$ decomposition\label{sec:conventions}}

\
 Given an $SO\left(9,1\right)$ chiral spinor $\lambda^{\alpha}$
(antichiral $\bar{\lambda}_{\alpha}$), one can write down its $SO\left(8\right)$
components through the use of the projectors $P_{I}^{\alpha}$ and
$\left(P_{I}^{\alpha}\right)^{-1}\equiv P_{\alpha}^{I}$, where $I$
generically indicates the $SO\left(8\right)$ indices, defined in
such a way that 
\begin{eqnarray*}
\lambda^{\alpha}=P_{a}^{\alpha}\lambda^{a}+P_{\dot{a}}^{\alpha}\lambda^{\dot{a}}, & \lambda^{a}=P_{\alpha}^{a}\lambda^{\alpha}, & \lambda^{\dot{a}}=P_{\alpha}^{\dot{a}}\lambda^{\alpha},\\
\bar{\lambda}_{\alpha}=P_{\alpha}^{a}\bar{\lambda}_{a}+P_{\alpha}^{\dot{a}}\bar{\lambda}_{\dot{a}}, & \bar{\lambda}_{a}=P_{a}^{\alpha}\bar{\lambda}_{\alpha}, & \bar{\lambda}_{\dot{a}}=P_{\dot{a}}^{\alpha}\bar{\lambda}_{\alpha}.
\end{eqnarray*}

Being invertible, they satisfy 
\[
\begin{array}{cc}
P_{a}^{\alpha}P_{\alpha}^{b}=\delta_{a}^{b}, & P_{\dot{a}}^{\alpha}P_{\alpha}^{\dot{b}}=\delta_{\dot{a}}^{\dot{b}},\\
P_{a}^{\alpha}P_{\alpha}^{\dot{a}}=0, & \delta_{\beta}^{\alpha}=P_{a}^{\alpha}P_{\beta}^{a}+P_{\dot{a}}^{\alpha}P_{\beta}^{\dot{a}},
\end{array}
\]
where $a,\dot{a}=1,\ldots,8$ are the $SO\left(8\right)$ spinorial
indices, representing different chiralities.

Note that upper and lower indices in the $SO\left(8\right)$ language
do not distinguish chiralities, \emph{i.e.}, one can define a spinorial
metric, $\eta_{ab}$ ($\eta_{\dot{a}\dot{b}}$), and its inverse,
$\eta^{ab}$ ($\eta^{\dot{a}\dot{b}}$), such that $\eta_{ac}\eta^{cb}=\delta_{a}^{b}$
($\eta_{\dot{a}\dot{c}}\eta^{\dot{c}\dot{b}}=\delta_{\dot{a}}^{\dot{b}}$)
and are responsible for lowering and raising spinorial indices, respectively,
acting as charge conjugation. For example, $\left(\sigma^{i}\right)^{\dot{a}a}=\eta^{ab}\eta^{\dot{a}\dot{b}}\left(\sigma^{i}\right)_{b\dot{b}}$.

Using the projectors, one can build a representation for the gamma
matrices $\gamma^{m}$ in terms of the $8$-dimensional equivalent
of the Pauli matrices,
\begin{equation}
\begin{array}{rclrcl}
\left(\gamma^{i}\right)^{\alpha\beta} & \equiv & \left(\sigma^{i}\right)^{\dot{a}a}\left(P_{\dot{a}}^{\alpha}P_{a}^{\beta}+P_{a}^{\alpha}P_{\dot{a}}^{\beta}\right), & \left(\gamma^{i}\right)_{\alpha\beta} & \equiv & \left(\sigma^{i}\right)_{a\dot{a}}\left(P_{\alpha}^{a}P_{\beta}^{\dot{a}}+P_{\alpha}^{\dot{a}}P_{\beta}^{a}\right),\\
\left(\gamma^{-}\right)^{\alpha\beta} & \equiv & \sqrt{2}\eta^{ab}P_{a}^{\alpha}P_{b}^{\beta}, & \left(\gamma^{-}\right)_{\alpha\beta} & \equiv & -\sqrt{2}\eta_{\dot{a}\dot{b}}P_{\alpha}^{\dot{a}}P_{\beta}^{\dot{b}},\\
\left(\gamma^{+}\right)^{\alpha\beta} & \equiv & \sqrt{2}\eta^{\dot{a}\dot{b}}P_{\dot{a}}^{\alpha}P_{\dot{b}}^{\beta}, & \left(\gamma^{+}\right)_{\alpha\beta} & \equiv & -\sqrt{2}\eta_{ab}P_{\alpha}^{a}P_{\beta}^{b},
\end{array}\label{eq:gammadecomposition}
\end{equation}
 where\begin{subequations} 
\begin{eqnarray}
\left(\sigma^{i}\right)_{a\dot{a}}\left(\sigma^{j}\right)^{\dot{a}b}+\left(\sigma^{j}\right)_{a\dot{a}}\left(\sigma^{i}\right)^{\dot{a}b} & = & 2\eta^{ij}\delta_{a}^{b},\\
\left(\sigma^{i}\right)^{\dot{a}a}\left(\sigma^{j}\right)_{a\dot{b}}+\left(\sigma^{j}\right)^{\dot{a}a}\left(\sigma^{i}\right)_{a\dot{b}} & = & 2\eta^{ij}\delta_{\dot{b}}^{\dot{a}},\\
\left(\sigma^{i}\right)_{a\dot{a}}\left(\sigma_{i}\right)_{c\dot{c}}+\left(\sigma^{i}\right)_{c\dot{a}}\left(\sigma_{i}\right)_{a\dot{c}} & = & 2\eta_{ac}\eta_{\dot{a}\dot{c}},\\
\left(\sigma^{ij}\right)_{\phantom{a}b}^{a}\left(\sigma^{ij}\right)_{\phantom{c}d}^{c} & = & 8\eta^{ac}\eta_{bd}-8\delta_{d}^{a}\delta_{b}^{c},\\
\left(\sigma^{ij}\right)_{\phantom{a}b}^{a}\left(\sigma^{ij}\right)_{\phantom{\dot{a}}\dot{b}}^{\dot{a}} & = & 4\sigma_{i}^{\dot{a}a}\sigma_{b\dot{b}}^{i}-4\delta_{b}^{a}\delta_{\dot{b}}^{\dot{a}},
\end{eqnarray}
\end{subequations}and $\eta^{ij}$ is the flat $SO\left(8\right)$
inverse metric, with $i,j=1,\ldots,8$. As usual, $\eta_{ik}\eta^{kj}=\delta_{i}^{j}$.
Note that 
\begin{equation}
\begin{array}{c}
\left\{ \gamma^{i},\gamma^{j}\right\} =2\eta^{ij},\\
\left\{ \gamma^{+},\gamma^{-}\right\} =-2,\\
\left\{ \gamma^{\pm},\gamma^{i}\right\} =\left\{ \gamma^{+},\gamma^{+}\right\} =\left\{ \gamma^{-},\gamma^{-}\right\} =0.
\end{array}
\end{equation}

Given a $SO\left(9,1\right)$ vector $N^{m}$, the $SO\left(8\right)$
decomposition used in this work is, 
\begin{equation}
N^{\pm}\equiv\frac{1}{\sqrt{2}}\left(N^{0}\pm N^{9}\right),
\end{equation}
and $N^{i}$, with $i=1,\ldots,8$, stands for the spatial components.
Therefore, the scalar product between $N^{m}$ e $P^{m}$ is given
by $N^{m}P_{m}=-N^{+}P^{-}-N^{-}P^{+}+N^{i}P_{i}$.

For a rank-$2$ antisymmetric tensor $N^{mn}$, the $SO\left(8\right)$
components will be represented as 
\[
\left\{ N^{ij},N^{i}=N^{-i},\bar{N}^{i}=N^{+i},N=N^{+-}\right\} .
\]

\end{document}